\documentstyle[12pt]{article}

\def \bsigma {\mbox {\boldmath $\sigma$}}
\def \bxi {\mbox {\boldmath $\xi$}}
\def \bpsi {\mbox {\boldmath $\psi$}}
\def \btheta {\mbox {\boldmath $\theta$}}
\def \bv {\mbox {\boldmath $v$}}
\def \bw {\mbox {\boldmath $w$}}
\def \bD {\mbox {\boldmath $D$}}

\def \bz {\mbox {\boldmath $z$}}

\begin{document}

\setcounter{page}{0}
\input epsf

\renewcommand{\baselinestretch}{1.0}

\title{\bf Mean-field dynamics of sequence processing neural
networks with finite connectivity}
\author{W. K. Theumann \footnote{e-mail: theumann@if.ufrgs.br}
\\Instituto de F\'{\i}sica,
Universidade Federal do Rio Grande do Sul,\\ Caixa Postal 15051.
91501-970 Porto Alegre, RS, Brazil.}

\date{\today}
\maketitle
\thispagestyle{empty}

\begin{abstract}

A recent dynamic mean-field theory for sequence processing in
fully connected neural networks of Hopfield-type (D\"uring, Coolen
and Sherrington, 1998) is extended and analyzed here for a
symmetrically diluted network with finite connectivity near
saturation. Equations for the dynamics and the stationary states
are obtained for the macroscopic observables and the precise
equivalence is established with the single-pattern retrieval
problem in a layered feed-forward network with finite
connectivity.

\end{abstract}

\vspace{2.0cm}

87.10.+e, 64.60.Cn, 07.05.Mh


\newpage
\setcounter{page}{1}

\section{Introduction}

A path-integral approach \cite{DD78,So87} has been successfully
applied already some time ago \cite{RSZ88} to study the dynamics
near saturation of the Hopfield model \cite{Ho82} with a symmetric
Hebbian learning rule that favors the retrieval of single
patterns. The results for the stationary states were found to be
those obtained by means of the replica method in equilibrium
statistical mechanics \cite{AGS87}. The performance of networks
trained with a sequence of stored patterns has been of interest
over some time $[6-13]$, for both biological and artificial neural
networks \cite{HKP91}, and the path-integral approach has been
applied to study the single-pattern stationary states in a 
mean-field dynamics near saturation for a fully connected recurrent 
network trained with a sequence of stored patterns \cite{DCS98}.

More recently, the transients in that network have been investigated
in a statistical neurodynamics approach, which was shown to be
equivalent to the path-integral method \cite{KO02}. These methods
are  particularly suitable to handle dynamical systems with
asymmetric interactions, as in the case of sequential patterns,
due to the absence of detailed balance where equilibrium statistical mechanics cannot be used.

Explicit equations were obtained for the retrieval overlap with
single patterns in a sequence, for the response function to an
external field and for the two-state correlation function \cite{DCS98,KO02}. By means of exact analytic work, supported by
numerical simulations, it has been found that the critical storage
capacity in the fully connected recurrent network for single-pattern retrieval in a sequence is $\alpha_c = 0.269$ \cite{DCS98}. This
is the same as the critical storage capacity for a different problem
in another network architecture, namely, the usual single-pattern
retrieval problem in the layered feed-forward network \cite{DKM89}.
It has been pointed out that this is due to the equivalence
between the two models in the absence of synaptic noise ($T=0$), and
that this equivalence does not extend beyond the $T=0$ limit. The
performance of the network turns out to be limited by a non-local
(in time) Gaussian noise in the effective single-spin problem
\cite{DCS98}.

The equilibrium behavior of Hopfield-type networks with finite
random decay (dilution) of synapses is of interest for biological
\cite{RTFP97} and artificial neural networks and has already been
considered for the single-pattern retrieval problem in systems
with or without microscopic detailed balance $[17,19-21]$. Dense
networks with connectivity $c=O(1)$ share most of the features of
fully connected networks and have a non-trivial dynamics, in
contrast to the extremely diluted network \cite{DGZ87}.

We discuss here the extension of the sequence processing dynamics
of D\"uring and al. to finite synaptic dilution and study the
precise relationship with the single-pattern dynamics and the
asymptotic stationary state of the layered feed-forward network.
This is a network that consists of identical layers of $N$
non-interacting units each, with synaptic interaction $J^{l}_{ij}$
between units $j$ and $i$ in layers $l$ and $l+1$, respectively.

The outline of the paper is the following. In Sec. 2 we present
the model and derive the connected generating functional. In Sec.
3 we discuss the effective single-site problem in dynamic
mean-field theory and present the specific equations for the
macroscopic observables, including the stationary state. We
discuss there the explicit relationship with the results for the
layered feed-forward network and summarize our conclusions in Sec. 4.

\section{The model}

\subsection{Defining relations}

We consider a network of $N$ Ising neurons in a microscopic state
$\bsigma(t)=\{\sigma_1(t),\dots,\sigma_N(t)\}$, at the time step
$t$ in which each $\sigma_i(t)=\pm 1$, that is updated
simultaneously according to the alignment of each unit to its local
field
\begin{equation}
h_i(t)=\sum_{j}J^{d}_{ij}\sigma_j(t) + \theta_i(t)\,\,\,,
\label{1}
\end{equation}
following a stochastic dynamics with transition probability
\begin{equation}
w_i[\sigma_i(t+1)|\sigma_i(t)]
=\frac{1}{2}[1+\sigma_i(t+1)\tanh(\beta \sigma_i(t)h_i(t))]\,\,\,,
\label{2}
\end{equation}
ruled by the noise-control parameter $\beta=T^{-1}$. The dynamics
is a deterministic one when $T=0$ and it is fully random when $T
=\infty$. In the former case, the state of a spin at a given time
is completely determined by the sign of its local field following
the relation $\sigma_i(t+1)={\rm sgn}[h_i(t)]$. Here, $J^{d}_{ij}$
is the synaptic coupling to other neurons and $\theta_i(t)$ is an
external field introduced in order to study the response of the
network to an external stimulus. Thus, given the probability
$p[\bsigma(0)]$ of an initial state, Eq.(2) determines the path
probability $p[{\bsigma}(0),\dots,{\bsigma}(t)]$ that takes the
system to the state $\bsigma(t)$ through a Markov chain.

A macroscopic sequential set of $p=\alpha N$ independent and
identically distributed random cyclic patterns $\bxi^{\mu}
=(\xi_1^{\mu},\dots,\xi_N^{\mu})$, $\mu=1,\dots,p$, such that
$\bxi^{p+1}=\bxi^{1}$ and each $\xi_i^{\mu}=\pm 1$ with
probability $\frac{1}{2}$, is assumed to be stored in the network
according to the learning rule
\begin{equation}
J^{d}_{ij}=\frac{c_{ij}}{cN}\sum_{\mu=1}^p\xi_i^{\mu+1}
\xi_j^{\mu}\,\,\,,
\label{3}
\end{equation}
and we are interested in the behavior of the network near
saturation, that is, for finite $\alpha$. The synaptic
interactions are assumed to be symmetrically diluted by means of
the set of couplings $\{c_{ij}\}$, with $c_{ij}=c_{ji}$ taken
from a set of independent identically distributed random
variables, which are the same for every time step, such that
$c_{ij}=1$ with probability $c$ and zero with probability $1-c$,
while $c_{ii}=0$. Thus, the average $[c_{ij}]=c$ becomes the
connectivity of the network and when $c=1$ one has the usual
synaptic coupling for the fully connected case. We consider a
dense network, with $c=O(1)$, and take $cN>>1$ allowing eventually
for a vanishingly small $c$ after taking the thermodynamic limit in
the final results of the dynamics. Once the network has a finite
connectivity it is expected to have a non-trivial dynamics which is
different from that of a purely random network as long as
$\beta\neq 0$. Thus, one has to ensure first a finite connectivity
and we do this next.

\subsection{The connected generating functional}

Since the matrix ${\bf J}^{d}=\{J^{d}_{ij}\}$ is non-symmetric,
one has to resort to a dynamical procedure to study the time 
evolution and the stationary states of the network and we follow previous
works using the generating functional or path-integral method
\cite{DCS98,KO02}. We introduce a connected generating
functional\\
\begin{equation}
Z_{c}(\bpsi)=\sum_{\bsigma(0),\dots,\bsigma(t)} \langle
p[{\bsigma}(0),\dots,{\bsigma}(t)]\rangle_{c_{ij}} e^{-i\sum_{s<t}
\bsigma(s).\bpsi(s)}\,\,\,,
\label{4}
\end{equation}
where $\langle \dots \rangle_{c_{ij}}$ denotes the path
probability for the connected network obtained by averaging
$p[{\bsigma}(0),\dots,{\bsigma}(t)]$ over the set $\{c_{ij}\}$
and, as usual, ${\bf x}.{\bf y} =\sum_{i}x_iy_i$. Here,
$\bpsi_i(s)=\{\psi(0),\dots,\psi(t-1)\}$ is introduced and
afterwards set to zero in order to generate state averages over
the path probability. Using an integral representation of unity
to account for the defining relation for the set of local fields
$\{h_i(s)\}$ given above, by means of an auxiliary set
$\{\hat{h}_i(s)\}$ \cite{DCS98}, we can separate the configurational
average over the set $\{c_{ij}\}$ and obtain
\begin{eqnarray}
Z_{c}(\bpsi) &=&\sum_{\bsigma(0),\dots,\bsigma(t)} p[\bsigma(0)]
\int\{{\rm d}{\bf h}{\rm d}{\bf \hat{h}}\}\nonumber\\
&&\prod_{s< t}\exp\{\beta \bsigma(s+1).{\bf h}(s)
-\sum_i \ln2\cosh({\beta}h_i(s))\nonumber\\
&+&i{\bf \hat{h}}(s).[{\bf h}(s)-\btheta(s)]-i\bpsi(s).
\bsigma(s)\}
\langle e^{-i{\bf \hat{h}}(s).{\bf J}^{d}\footnotesize{\bsigma}(s)}
\rangle_{c_{ij}}\,,
\label{5}
\end{eqnarray}
where $\{{\rm d}{\bf h}{\rm d}{\bf \hat{h}}\}
=\prod_i\prod_{s<t}[{\rm d}h_i(s){\rm d}\hat{h}_i(s)/2\pi]$.

We proceed next to calculate the configurational average along similar
lines to those used in statistical mechanics of diluted spin-glass
and neural network models \cite{VB85,TE01} and write
\begin{eqnarray}
\left\langle e^{-i\hat{{\bf h}}(s).{\bf J}^{d}\footnotesize{\bsigma}(s)}
\right\rangle_{c_{ij}}&=&\prod_{i,j}\{1+c(\exp[-i\hat{h_i}(s)\tilde{J}_{ij}
\sigma_j(s)]-1)\}\nonumber\\
&=&\exp[\sum_{i,j}\ln (1+c\eta_{i,j}(s))]\,\,\,, \label{6}
\end{eqnarray}
where the coupling $\tilde{J}_{ij}= J^{d}_{ij}$ for $c_{ij}=1$ and
$\eta_{i,j}(s)\equiv\exp[-i\hat{h_i}(s)\tilde{J}_{ij}\sigma_j(s)]
-1$. Keeping in mind that we want results for finite $c$ and
noting that $\tilde{J}_{ij}=O(\sqrt{\alpha/cN})$ is small, we use
first $\eta_{i,j}$ as an expansion parameter up to second order
and then expand in $\tilde{J}_{ij}$, to that order, and obtain
\begin{equation}
\langle e^{-i\hat{{\bf h}}(s).{\bf J}^{d}\footnotesize{\bsigma}(s)}
\rangle_{c_{ij}}=\exp\{-i\hat{{\bf h}}(s).{\bf J}
{\bf \bsigma}(s)
-\frac{1}{2}c(1-c)\sum_{i,j}[\hat{h_i}(s)\tilde{J}_{ij}
\sigma_j(s)]^2\}\,\,\,.
\label{7}
\end{equation}
The couplings on the right ${\bf J}=\{J_{ij}\}$, with
$J_{ij}=N^{-1}\sum_{\mu}\xi_i^{\mu+1}\xi_j^{\mu}$, are now the
synapses for the fully connected network with sequential patterns
and the second term in the argument of the exponential contains
the full effect of the dilution.

Before doing the average of the exponential over the patterns
following the standard procedure below, one may replace here the
pattern dependence of the last term, up to terms of higher order
which are irrelevant in mean-field theory, by its average over the
pattern distribution becoming thereby pattern independent. In this
way, the state $\bsigma(s)$ becomes decoupled from the auxiliary
field $\hat{{\bf h}}(s)$ and, noting that $\sum_i\sigma_{i}^2(s)=N$,
we obtain
\begin{equation}
\left\langle e^{-i\hat{{\bf h}}(s).{\bf J}\footnotesize{\bsigma}(s)}
\right\rangle_{c_{ij}}=e^{-i\hat{{\bf h}}(s).{\bf J}
\footnotesize{\bsigma}(s)}
e^{-\Delta^{2}{\bf \hat{h}}^{2}(s)/2}\,\,\,,
\label{8}
\end{equation}
where $\Delta^2=\alpha(1-c)/c$ and $\hat{{\bf h}}^{2}(s)=
\sum_i\hat{h}_{i}^2(s)$. Note that $J_{ij}$ is of order
$1/\sqrt{N}$, the local field $h_{i}$ and $\hat{h}_{i}$ are of
order one and, consequently, the argument of both exponents is, as
one would expect, of order $N$. Incidentally, the result in Eq.(8)
is precisely the average $\langle\exp(-i\hat{{\bf h}}(s).{\bf
J}^{eff} {\bf \bsigma}(s))\rangle_{\{\delta_{ij}\}}$ over the
pattern-independent set $\{\delta_{ij}\}$ of independent Gaussian
random variables with mean zero and variance $\Delta^2$, in
which the effective coupling $J^{eff}_{ij}=J_{ij}+\delta_{ij}$.
Thus, as in the case of the equilibrium behavior of networks, the
finite symmetric dilution in the dynamics amounts to the addition
of a static Gaussian noise to the synaptic couplings of the fully 
connected network. \cite{HS86}.

This completes the derivation of the formal expression for
$Z_{c}(\bpsi)$, in which the connected path probability takes
the form
\begin{eqnarray}
\langle p[{\bsigma}(0),\dots,{\bsigma}(t)]\rangle_{c_{ij}}
&=&p[\bsigma(0)]\int\{{\rm d}{\bf h}{\rm d}{\bf \hat{h}}\}
\prod_{s< t}\exp\{\beta \bsigma(s+1).{\bf h}(s)\nonumber\\
&-&\sum_i \ln2\cosh({\beta}h_i(s))
+i{\bf \hat{h}}(s).[{\bf h}(s)-\btheta(s)]\nonumber\\
&-&i{\bf \hat{h}}(s). {\bf J}\bsigma(s)
-\frac{1}{2}\Delta^{2}{\bf \hat{h}}^{2}(s)\}\,\,\,.
\label{9}
\end{eqnarray}
Clearly, the auxiliary variable ${\bf \hat{h}}$ has the role of
relating microscopic physical quantities with each other and this
will show up next in constructing the macroscopic observables.

\subsection{The macroscopic observables}

The macroscopic dynamic observables depend on the problem one
wants to study. If one is interested in the retrieval of a single
pattern, say $\xi_i^{s}$ at each time step $s$, the case to which
we restrict ourselves in this work, one assumes that that pattern
is `condensed' which means that the overlap with the state of the
network, $\bsigma(s)$, at that time step is of $O(1)$ and that all
other overlaps are of $O(N^{-\frac{1}{2}})$. The non-condensed
overlaps introduce a stochastic noise into the dynamics which may
be thought as a quenched disorder. One then expects that the
relevant macroscopic variables for single-pattern retrieval are
the overlap $m_N(s)$, the single-site response and correlation
functions, $G_N(s,s\prime)$ and $C_N(s,s\prime)$, respectively,
defined and given, for large but finite $N$, by
\begin{equation}
m^{s}_N(s)\equiv\frac{1}{N}\sum_i\xi_i^{s}\langle\sigma_i(s)\rangle
=i\lim_{\footnotesize{\bpsi}\rightarrow
0}\frac{1}{N}\sum_i\xi_i^{s} \frac{\partial Z_{c}(\bpsi)}{\partial
\psi_i(s)}\,\,\,, \label{10}
\end{equation}
\begin{equation}
G_N(s,s\prime)\equiv\frac{1}{N}\sum_i \frac{\partial
\langle\sigma_i(s)\rangle}{\partial \theta_i(s\prime)}
=i\lim_{\footnotesize{\bpsi}\rightarrow 0}\frac{1}{N}\sum_i
\frac{\partial^{2}Z_{c}(\bpsi)}{\partial \psi_i(s)\partial
\theta_i(s\prime)}\,\,\,,
\label{11}
\end{equation}
\begin{equation}
C_N(s,s\prime)\equiv\frac{1}{N}\sum_i
\langle\sigma_i(s)\sigma_i(s\prime)\rangle\nonumber\\
=-\lim_{\footnotesize{\bpsi}\rightarrow 0}\frac{1}{N}\sum_i
\frac{\partial^{2}Z_{c}(\bpsi)}{\partial \psi_i(s)
\partial \psi_i(s\prime)}\,\,\,,
\label{12}
\end{equation}\\
where the brackets now denote the thermal averages over the
connected path probability given by Eq.(9). Note that, due to
the form of $Z_{c}(\bpsi)$, Eq.(5), the response function is
given by the $\bpsi(s)\rightarrow 0$ limit of
$-iN^{-1}\sum_i\langle\sigma_i(s)\hat{h}_i(s\prime)\rangle$. In
principle, there could be other observables, as $k_N(s)
=N^{-1}\sum_i\xi_i^{s+1}\hat{h}_i(s)$ and $Q_N(s,s\prime)=
N^{-1}\sum_i\hat{h}_i(s)\hat{h}_i(s\prime)$, but they do not
represent physical quantities so that one would expect that they
do not show up in mean-field theory. Indeed, it has been found that
the saddle-point solutions for these observables turn out to vanish
in mean-field theory \cite{DCS98}.

\section{Dynamic mean-field theory}

As usual in neural network studies, one can only deal with a
finite number of condensed patterns. Having assumed that at each
time-step a single pattern is condensed, a finite sequence of
patterns with $\mu\leq t$ will be condensed in fixed time $t$.
Thus, the dynamical mean-field theory will be restricted to finite
time scales although we keep the limit $N\rightarrow\infty$. In
that limit, the theory should be self-averaging and we are then left 
with an explicit average of the generating functional over the 
non-condensed patterns, $\overline {Z_{c}(\bpsi)]}$. Noting that 
only the part of the exponential depending on ${\bf J}$ on the right 
in Eq.(9) is pattern dependent, one can follow the steps in ref. 15
to do this average and the relevant macroscopic observables in the
dynamic mean-field theory, in the limit $N\rightarrow\infty$,
become
\begin{equation}
m^s(s)=\lim_{N\rightarrow\infty}\frac{1}{N}\sum_i
\overline{\xi_i^{s}\langle\sigma_i(s)\rangle}
=i\lim_{\footnotesize{\bpsi}\rightarrow
0}\lim_{N\rightarrow\infty} \frac{1}{N}\sum_i\xi_i^{s}
\frac{\partial\overline{Z_{c}(\bpsi)}}{\partial \psi_i(s)}\,\,\,,
\label{13}
\end{equation}
\begin{eqnarray}
G(s,s\prime)=-i\lim_{N\rightarrow\infty}\frac{1}{N}\sum_i
\overline{\langle\sigma_i(s)\hat{h}_i(s\prime)\rangle}
=i\lim_{\footnotesize{\bpsi}\rightarrow
0}\lim_{N\rightarrow\infty} \frac{1}{N}\sum_i\frac{\partial^{2}
\overline{Z_{c}(\bpsi)}}{\partial \psi_i(s)\partial
\theta_i(s\prime)}\,\,\,,
\label{14}
\end{eqnarray}
\begin{eqnarray}
C(s,s\prime)=\lim_{N\rightarrow\infty}\frac{1}{N}\sum_i
\overline{\langle\sigma_i(s)\sigma_i(s\prime)\rangle}
=-\lim_{\footnotesize{\bpsi}\rightarrow
0}\lim_{N\rightarrow\infty}
\frac{1}{N}\sum_i\frac{\partial^{2}\overline{Z_{c}(\bpsi)}}
{\partial \psi_i(s)\partial \psi_i(s\prime)}\,\,\,.
\label{15}
\end{eqnarray}

\subsection{Single-site relations}

An explicit expression is then obtained in the form of an
effective single-site normalized average for any function
$f[\{\sigma\}]$ of the states of the diluted network with
finite connectivity, denoted as $\langle
f[\{\sigma\}]\rangle^{*}$, for $p=\alpha N$ in the limit
$N\rightarrow\infty$. The normalization is such that the average
is unity when $f[\{\sigma\}]=1$, and the result becomes
independent of the specific site after the gauge transformation
$\sigma(s) \rightarrow\sigma(s)\xi_i^{s}$ and $h(s)\rightarrow
h(s)\xi_i^{s+1}$ \cite{DCS98}. This expression reads,
\begin{eqnarray}
\langle
f[\{\sigma\}]\rangle^{*}&=&\sum_{\sigma(0),\dots,\sigma(t)}
\int\{{\rm d}h{\rm d}\hat{h}\}p(\sigma(0))f[\{\sigma\}]
\exp\{\sum_{s< t}[\beta \sigma(s+1)h(s)\nonumber\\
&-&\ln2\cosh(\beta h(s))]+i\sum_{s<t}\hat{h}(s)[h(s)
-\theta(s)-m^s(s)]\nonumber\\
&-&\frac{1}{2}\alpha \sum_{s,s\prime<t}\hat{h}(s)
D(s,s\prime)\hat{h}(s\prime)\}\,\,\,,
\label{16}
\end{eqnarray}
in which $\{{\rm d}h{\rm d}\hat{h}\}
=\prod_{s<t}[{\rm d}h(s){\rm d}\hat{h}(s)/2\pi]$ and all 
quantities are single-site quantities. Here,
\begin{equation}
D(s,s\prime)=R(s,s\prime)+\frac{1-c}{c}\,\,\delta_{s,s\prime}\,\,\,,
\label{17}
\end{equation}
is the covariance $D(s,s\prime)=\langle v(s)v(s\prime)\rangle$ of
the zero-average Gaussian random noise $v(s)$ in the effective local
field
\begin{equation}
h(s)=m^s(s)+\theta(s)+\sqrt{\alpha }v(s)\,\,\,, \label{18}
\end{equation}
as can be seen readily by integration over $\hat h$. Thus, the
effect of the dilution appears as a local term (in time)
contributing to the variance $\langle v^{2}(s)\rangle$ of the noise.
It will be seen below that, precisely because of its
local character, the dilution term has a crucial role in determining
the stationary state of the network. The non-local contribution is
that in the covariance of the noise, $R(s,s\prime)$, for the fully
connected network and this is given by the recurrence relation
\cite{KO02}
\begin{equation}
R(s,s\prime)=C(s,s\prime)+G(s,s-1)G(s\prime,s\prime-1)
R(s-1,s\prime-1)\,\,\,. \label{19}
\end{equation}
Combining Eqs.(17) and (19) with the appropriate average over the
states given by Eq.(16), one obtains the evolution equations for
the macroscopic observables.

\subsection{Equations for the dynamics}

The single-site equations for the time evolution of the macroscopic
observables are then given by
\begin{eqnarray}
m(s)&=&\int\{{\rm d}{\bv}{\rm d}{\bw}\}\,e^{i{\bv}.{\bw}
-\frac{1}{2}{\bw.\bD\bw}}\nonumber\\
&&\times \tanh\beta[m(s-1)+\theta(s-1) +\sqrt{\alpha}v(s-1)]\,\,,
\label{20}
\end{eqnarray}
\begin{eqnarray}
G(s,s\prime)&=&\beta\delta_{s,s\prime+1}\{1-
\int\{{\rm d}{\bv}{\rm d}{\bw}\}\,e^{i{\bv.\bw}
-\frac{1}{2}{\bw.\bD\bw}}\nonumber\\
&&\times \tanh^{2}\beta[m(s-1)+\theta(s-1)+\sqrt{\alpha}v(s-1)]\}
\,\,, \label{21}
\end{eqnarray}
\begin{eqnarray}
C(s,s\prime)&=&\delta_{s,s\prime}+(1-\delta_{s,s\prime})
\int\{{\rm d}{\bv}{\rm d}{\bw}\}\,e^{i{\bv.\bw}
-\frac{1}{2}{\bw.\bD\bw}}\nonumber\\
&&\times
\tanh\beta[m(s-1)+\theta(s-1)+\sqrt{\alpha}v(s-1)]\nonumber\\
&&\times \tanh\beta[m(s\prime-1)+\theta(s\prime-1) +\sqrt{\alpha
}v(s\prime-1)]\,\,, \label{22}
\end{eqnarray}
where $\bD$ is the covariance matrix of elements $D(s,s\prime)$.
Note that the arguments of the $\tanh$ are $\beta$ times the local
field at time $s$ or $s\prime$. From here on we drop the pattern
index in the overlap $m^s(s)$ with the understanding that $m(s)$
means the overlap with the pattern in the sequence that
corresponds to that time step.

It has been pointed out before, for the fully connected network
\cite{DCS98}, that the response function differs from zero only
for changes of state one time step after the perturbation in the
external field and that, consequently, stationary macroscopic
states should be reached in finite timescales. Clearly, that is 
still the case for the diluted network.

The equations can be reduced by explicit integration over $\bw$,
noting that only the matrix element $D(s-1,s-1)$ is needed for the
first two of them and we obtain
\begin{equation}
m(s)=\int Dz \,\tanh\beta[m(s-1)+\theta(s-1)+z\sqrt{\alpha
D(s-1,s-1)}]\,\,\,, \label{23}
\end{equation}
\begin{eqnarray}
G(s,s-1)&=&\beta\{1-\int Dz \nonumber\\
&&\times\tanh^{2}\beta[m(s-1) +\theta(s-1)+z\sqrt{\alpha
D(s-1,s-1)}]\}\,\,\,, \label{24}
\end{eqnarray}
where $z$ is a Gaussian random variable with mean zero and unit
variance and, as usual, $Dz=\frac{1}{\sqrt{2\pi}}e^{-z^{2}/2}$. The
third equation becomes
\begin{eqnarray}
C(s,s\prime)&=&(2\pi|\bD_{11}|)^{-1/2}
\int{\rm d}\bz\, e^{-\frac{1}{2}\bz.\bD^{-1}_{11}\bz}\nonumber\\
&&\times\tanh\beta[m(s-1)+\theta(s-1)+\sqrt{\alpha}z(s-1)]\nonumber\\
&&\times\tanh\beta[m(s\prime-1)+\theta(s\prime-1) +\sqrt{\alpha
}z(s\prime-1)]\,\,\,, \label{25}
\end{eqnarray}
when $s\neq s\prime$ and $C(s,s)=1$. Here, $\bD_{11}$ is the $2$ x $2$ submatrix of $\bD$ involving the elements at times $s-1$ and $s\prime-1$, 
while $\bz$ is the vector with components $z(s-1)$ and $z(s\prime-1)$. The
diagonal elements of $\bD_{11}$ are given by
\begin{eqnarray}
D(s,s)&=&R(s,s)+(1-c)/c\nonumber\\
R(s,s)&=&1+G^{2}(s,s-1)R(s-1,s-1) \label{26}
\end{eqnarray}
and only the off-diagonal elements depend on the correlation function,  
which enters the dynamics through Eq.(25).

In the case where $c=1$, our equations, coincide with those for
the transients in the fully connected network \cite{KO02}. In the
extremely dilute limit $c\rightarrow 0$, the storage ratio
$\alpha$ has to be replaced by the ratio $p/cN$, that is, the
number of patterns per connected units, which means $\alpha/c$.
The equations for the overlap and the response function become
then those for the extremely diluted network, and the correlation 
function vanishes in that limit. Given an initial value for $R(s,s)$, Eqs.(23),(24) and (26) form a closed set of equations for the time 
evolution of the macroscopic observables of primary interest. It is
important to keep in mind that these quantities do not depend on the
correlation function.

It may be noted at this point that, for general $c$, Eqs.(23),(24)
and (26) are formally the same as the recurrence relations for the
overlap and the stochastic noise in the layered feed-forward
network with finite connectivity for the single-pattern retrieval
problem \cite{DKM89}. The task in this problem is to recognize a
given single pattern as the states evolve from one layer to the 
next one. The network model consists of $N$ binary units placed on 
layers in which all units of a layer are updated simultaneously 
according to the alignment with their local field produced by the 
units in the previous layer and there is no feedback from any layer 
to previous ones. A set of independent random patterns, $\{\xi_1^{\mu}(l),\dots,\xi_N^{\mu}(l)\}$,
$\mu=1,\dots,p$, each $\xi_i^{\mu}(l)=\pm 1$ with probability
$\frac{1}{2}$, is generated on every layer $l$ independently of
other layers, and the patterns are assumed to be stored in the
network according to the learning rule
\begin{equation}
J^{d}_{ij}(l)=\frac{c_{ij}}{cN}\sum_{\mu=1}^p\xi_i^{\mu}(l+1)
\xi_j^{\mu}(l)\,\,\,. \label{27}
\end{equation}

The local field at unit $i$ on layer $l$ may then be
written as
\begin{equation}
h_i(l)=\xi_i^{1}(l+1)m^{1}(l)+\theta_i(l)+z_i(l)\,\,\,, \label{28}
\end{equation}
where the first term is the signal from the finite overlap with 
pattern $\bxi^1$, say,
$m^{1}(l)$
and
\begin{equation}
z_i(l)=\sum_{\mu\neq 1}\xi_i^{\mu}(l+1)m^{\mu}(l) \label{29}
\end{equation}
is the noise due to the remaining patterns with overlap
$m^{\mu}(l)$. The recurrence relations for the overlap $m^{1}(l)$
and for the variance of the noise, $D(l,l)$, are then given 
precisely by Eqs.(23), (24) and (26) if the time step is replaced 
by the layer index and writing $G(l,l-1)=\beta[1-\tilde q(l-1))]$, 
where \cite{DKM89}
\begin{equation}
\tilde q(l-1)=\int Dz\,\tanh^{2}\beta[m(l-1)+\theta(l-1)
+z\sqrt{\alpha D(l-1,l-1)}] \label{30}
\end{equation}
with
\begin{equation}
R(l,l)=1+\beta^{2}[1-\tilde q(l-1)]^{2}R(l-1,l-1) \label{31}
\end{equation}
and the relationship with $D(l,l)$ given above. The variable
$\tilde q(l)$ is, clearly, the single-site average
$\overline{\langle\sigma_i(l)\rangle^2}$ in the layered network,
where the brackets denote here the average with the distribution
given by Eq.(2).


Consider now the correlation function in the layered network. To
obtain $C(l,l\prime)$ one needs to establish a recurrence relation
for the covariance $\overline{\langle z_i(l)z_i(l\prime)\rangle}$
of the noise. Following a standard procedure \cite{KO02,DKM89},
one finds that, for $\alpha\neq 0$, the covariance is zero due to
the independence of the patterns in different layers and,
consequently, $C(l,l\prime)=\delta_{l,l\prime}$.

\subsection{The stationary state}

The stationary state of the network is obtained from the
time-translation invariant limit cycle solutions
\begin{equation}
m(s)=m\,\,,\,\, C(s,s\prime)=C(s-s\prime)\,\,,\,\, G(s,s\prime)
=G(s-s\prime) \label{32}
\end{equation}
for the macroscopic observables. Then, $R(s,s\prime)=R(s-s\prime)$
and $D(s,s\prime)=D(s-s\prime)$ with $R(s,s\prime)$ given by
Eq.(19). The equations for the first two macroscopic observables
in the stationary state become then
\begin{equation}
m=\int Dz \,\tanh\beta[m+\theta+z\sqrt{\alpha
(\rho+(1-c)/c)}]\,\,\,, \label{33}
\end{equation}
\begin{equation}
\tilde q=\int Dz \tanh^{2}\beta[m +\theta+z\sqrt{\alpha
(\rho+(1-c)/c)}]\,\,\,, \label{34}
\end{equation}
where $\tilde q$ is given by $G(1)=\beta(1-\tilde q)$. Here,
\begin{equation}
\rho=\frac{1}{1-\beta^{2}(1-\tilde q)^{2}}\,\,\,, \label{35}
\end{equation}
is the stationary value of the variance in the internal
field for the fully connected network. These are, again, the
equations for the stationary state for the single-pattern
retrieval problem in the layered feed-forward network with
finite connectivity and with an effective Gaussian noise of
variance $\tilde \rho=\alpha(\rho+(1-c)/c)$. The results at $T=0$
have already been discussed, and it has been found that both,
the critical storage capacity $\alpha_{c}$ and the limiting overlap
of the recalled pattern, are monotonic decreasing functions of the
dilution \cite{DKM89}.

To obtain the correlation function in the stationary state it is
convenient to separate in Eq.(22) the persistent and non-persistent
parts in $R(\tau)$ \cite{DCS98},
\begin{equation}
R(\tau)=r+\tilde R(\tau)\,\,,\,\,
\lim_{\tau\rightarrow\infty}\tilde R(\tau)=0\,\,\,, \label{36}
\end{equation}
which yields $D(\tau)=r+\tilde D(\tau)$, where
\begin{equation}
\tilde D(\tau) =\tilde R(\tau)+[(1-c)/c]\,\delta_{\tau,0}
\label{37}
\end{equation}
also vanishes in that limit. The separation of the persistent part
in $D(\tau)$ and a linearization introduces a new Gaussian
integration over a variable $z$ of mean zero and variance
$\alpha r$, where $r=q\rho$ and $q
=\lim_{\tau\rightarrow\infty} C(\tau)$ is the persistent part of
$C(\tau)$. This may be interpreted as the spin-glass order parameter. 
The remaining Gaussian random variable has covariance
$\alpha \tilde D(\tau)$ and, in the limit $\tau\rightarrow\infty$,
$v(\tau)$ and $v(0)$ become uncorrelated. We are left with the
variance $\alpha \tilde D(0)$, where
\begin{equation}
\tilde D(0)=(1-q)\rho+\frac{1-c}{c}\,\,, \label{38}
\end{equation}
and we obtain for the persistent part of the correlation
function
\begin{equation}
q=\int Dz\,\{\int Dx\, \tanh\beta[m+\theta+z\sqrt{\alpha q\rho}
+x\sqrt{\alpha((1-q)\rho+(1-c)/c)}]\}^{2}. \label{39}
\end{equation}

We consider next the solutions of this equation combined with the
response function in the absence of recall for $\theta=0$. This
requires $\tilde q$, which is always non-zero and vanishes at high
$T$ as
\begin{equation}
\tilde q=\alpha\beta^2(1+\frac{1-c}{c})+O(\beta^4)\,\,\,.
\label{40}
\end{equation}
In the limit $T\rightarrow 0$, $\beta(1-\tilde q)$ remains finite
and this leads to $\rho=1+2/\pi\alpha$ when $m=0$. Consider next
the equation for $q$ and note that it always has the paramagnetic
solution $q=0$ in the absence of recall. In the limit
$T\rightarrow 0$, we find
\begin{equation}
q=\int Dz\, {\rm erf}^{2}[\frac{z\sqrt{q\rho}}
{\sqrt{2[(1-q)\rho+(1-c)/c]}}]\,\,\,, \label{41}
\end{equation}\\
for non-zero $\alpha$, and analysis of this equation shows that,
for any connectivity $c<1$, the only solution is $q=0$. In
contrast, there is also the solution $q=1$ in the case of the
fully connected network. We also conclude that there
is no solution other than $q=0$ for finite $T$, regardless of the
connectivity, in extension of earlier work \cite{DCS98}. Thus, the
persistent part of the correlation function vanishes for all $T$
in the sequence processing network when $c<1$. In contrast, it
is always zero in the layered network, as we saw above.

\section{Summary and conclusions}

We extended the path-integral method to study the retrieval dynamics
of consecutive single patterns in a sequence for a symmetrically
diluted network near saturation in a dynamic mean-field theory. This
is an exact procedure for an asymptotically large network and the
close agreement with numerical simulations obtained in recent works
\cite{DCS98,KO02} seem to indicate that the theory accounts for the
essentials of single-sequence processing. The main feature is the
presence of an, in general, non-local Gaussian stochastic noise in
time in the effective single-site problem. Only the local part of
that noise enters in the equations for the dynamics and the
stationary states for the overlap with a pattern and for the response
function to an external field.

The symmetric dilution of the synapses, which is equivalent to the
addition of a static noise to the coupling of the fully connected
network, introduces a further local Gaussian noise which is relevant
for the retrieval performance of the network. We have shown that the
resulting equations for the dynamics and for the stationary states
of the sequence processing network are precisely the same as those
for the single-pattern retrieval problem in the layered feed-forward
network, for all $T$. We also showed that the particular form of the
resulting Gaussian stochastic noise is exactly the same in both systems
for any finite connectivity.

Nevertheless, the two systems are not fully equivalent. Indeed,
although the two-time correlation function is zero in the layered
network, there is a finite, persistent correlation $q=1$ for the
fully connected sequence processing network at $T=0$ \cite{DCS98}.
However, we also showed that this finite value disappears as soon
as there is a dilution of the couplings, restoring in that case
the formal equivalence between the performance of both systems. We
insist here that the task of the network and its architecture are
different for both. In the case of finite $T$ there is no
persistent part of the two-time correlation function in both
systems, regardless of the connectivity.

Finally, all the results discussed here with the path-integral
method also follow from the statistical neurodynamics approach. It
would be interesting to study the effects of synaptic dilution on
other tasks in a sequence processing network, like mixture-state
retrieval, in which the overlap with more that one pattern in a
sequence remains finite. It may also be interesting to introduce a
dynamical dilution process that could account for deteriorating
synapses. The random symmetric dilution used here is a static one
that is not altered in the evolution of the network. These are issues
that will be studied in future work.\\

\thanks{\bf Acknowledgments}\\

We thank Alba Theumann for discussions at an early stage of the work. This work was financially supported in part by CNPq (Conselho Nacional de Desenvolvimento Cient{\'\i}fico e Tecnol{\'o}gico), Brazil.

\end{document}